\documentclass[aps,pra,floatfix,tightenlines,superscriptaddress,twocolumn,showpacs,amsmath,amssymb]{revtex4}
\usepackage{times}

\usepackage{graphicx}
\usepackage{color}
\usepackage{dcolumn}% Align table columns on decimal point
\usepackage{bm}% bold math
\begin{document}

% \preprint{APS/123-QED}

\title{
   Fast Quantum Modular Exponentiation
}

\author{Rodney Van Meter}
\email{rdv@tera.ics.keio.ac.jp}
\author{Kohei M. Itoh}%
% \email{kitoh@appi.keio.ac.jp}
\affiliation{%
  Graduate School of Science and Technology, Keio University and CREST-JST\\
 3-14-1 Hiyoushi, Kohoku-ku, Yokohama-shi, Kanagawa 223-8522, Japan
}%

% \date{\today}
\date{submitted July 28, 2004; revised Feb. 9, 2005; accepted Mar. 4,
  2005.}

\begin{abstract}
We present a detailed analysis of the impact on quantum modular
exponentiation of architectural features and possible concurrent gate
execution.  Various arithmetic algorithms are evaluated for execution
time, potential concurrency, and space tradeoffs.  We find that to
exponentiate an $n$-bit number, for storage space $100n$ (twenty times
the minimum $5n$), we can execute modular exponentiation two hundred
to seven hundred times faster than optimized versions of the basic
algorithms, depending on architecture, for $n=128$.  Addition on a
neighbor-only architecture is limited to $O(n)$ time while
non-neighbor architectures can reach $O(\log{n})$, demonstrating that
physical characteristics of a computing device have an important
impact on both real-world running time and asymptotic behavior.  Our
results will help guide experimental implementations of quantum
algorithms and devices.
\end{abstract}


\maketitle
% \bibliographystyle{abbrv}
%\pagebreak
%\newpage
%\end{titlepage}

%########### body 

\section{Introduction}

Research in quantum computing is motivated by the possibility of
enormous gains in computational
time~\cite{shor:siam-factor,grover96,deutsch-jozsa92,nielsen-chuang:qci}.
The process of writing programs for quantum computers naturally
depends on the architecture, but the application of classical computer
architecture principles to the architecture of quantum computers
has only just begun.

Shor's algorithm for factoring large numbers in polynomial time is
perhaps the most famous result to date in the
field~\cite{shor:siam-factor}.  Since this algorithm is well defined
and important, we will use it as an example to examine the
relationship between architecture and program efficiency, especially
parallel execution of quantum algorithms.  Shor's factoring algorithm
consists of main two parts, quantum modular exponentiation, followed
by the quantum Fourier transform.  In this paper we will concentrate
on the quantum modular exponentiation, both because it is the most
computationally intensive part of the algorithm, and because
arithmetic circuits are fundamental building blocks we expect to be
useful for many algorithms.

Fundamentally, quantum modular exponentiation is $O(n^3)$; that is,
the number of quantum gates or operations scales with the cube of the
length in bits of the number to be
factored~\cite{shor:factor,vedral:quant-arith,beckman96:eff-net-quant-fact}.
It consists of $2n$ modular multiplications, each of which consists of
$O(n)$ additions, each of which requires $O(n)$ operations.  However,
$O(n^3)$ {\em operations} do not necessarily require $O(n^3)$ {\em
time steps}.  On an abstract machine, it is relatively straightforward
to see how to reduce each of those three layers to $O(\log n)$ time
steps, in exchange for more space and more {\em total} gates, giving a
total running time of $O(\log^3 n)$ if $O(n^3)$ qubits are available
and an arbitrary number of gates can be executed concurrently on
separate qubits.  Such large numbers of qubits are not expected to be
practical for the foreseeable future, so much interesting engineering
lies in optimizing for a given set of constraints.  This paper
quantitatively explores those tradeoffs.

This paper is intended to help guide the design and experimental
implementation of actual quantum computing devices as the number of
qubits grows over the next several generations of devices.  Depending
on the post-quantum error correction, application-level effective
clock rate for a specific technology, choice of exponentiation
algorithm may be the difference between hours of computation time and
weeks, or between seconds and hours.  This difference, in turn,
feeds back into the system requirements for the necessary strength of
error correction and coherence time.

The Sch\"onhage-Strassen multiplication algorithm is often quoted in
quantum computing research as being $O(n \log n \log \log n)$ for a
single multiplication~\cite{knuth:v2-3rd}.  However, simply citing
Sch\"onhage-Strassen without further qualification is misleading for
several reasons.  Most importantly, the constant factors
matter~\footnote{Shor noted this in his original paper, without
explicitly specifying a bound.  Note also that this bound is for a
Turing machine; a random-access machine can reach $O(n\log n)$.}:
quantum modular exponentiation based on Sch\"onhage-Strassen is only
faster than basic $O(n^3)$ algorithms for more than approximately 32
{\em kilobits}.  In this paper, we will concentrate on smaller problem
sizes, and exact, rather than $O(\cdot)$, performance.

Concurrent quantum computation is the execution of more than one
quantum gate on independent qubits at the same time.  Utilizing
concurrency, the latency, or circuit depth, to execute a number of
gates can be smaller than the number itself.  Circuit depth is
explicitly considered in Cleve and Watrous' parallel implementation of
the quantum Fourier transform~\cite{cleve:qft}, Gossett's quantum
carry-save arithmetic~\cite{gossett98:q-carry-save}, and Zalka's
Sch\"onhage-Strassen-based implementation~\cite{zalka98:_fast_shor}.
Moore and Nilsson define the computational complexity class {\bf QNC}
to describe certain parallelizable circuits, and show which gates can
be performed concurrently, proving that any circuit composed
exclusively of Control-NOTs (CNOTs) can be parallelized to be of depth
$O(\log n)$ using $O(n^2)$ ancillae on an abstract
machine~\cite{moore98:_parallel_quantum}.

We analyze two separate architectures, still abstract but with some
important features that help us understand performance.  For both
architectures, we assume any qubit can be the control or target for
only one gate at a time.  The first, the $AC$, or {\em Abstract
Concurrent}, architecture, is our abstract model.  It supports CCNOT
(the three-qubit Toffoli gate, or Control-Control-NOT), arbitrary
concurrency, and gate operands any distance apart without penalty.  It
does not support arbitrary control strings on control operations, only
CCNOT with two ones as control.  The second, the $NTC$, or {\em
Neighbor-only, Two-qubit-gate, Concurrent} architecture, is similar
but does not support CCNOT, only two-qubit gates, and assumes the
qubits are laid out in a one-dimensional line, and only neighboring
qubits can interact.  The 1D layout will have the highest
communications costs among possible physical topologies.  Most real,
scalable architectures will have constraints with this flavor, if
different details, so $AC$ and $NTC$ can be viewed as bounds within
which many real architectures will fall.  The layout of variables on
this structure has a large impact on performance; what is presented
here is the best we have discovered to date, but we do not claim it is
optimal.

The $NTC$ model is a reasonable description of several important
experimental approaches, including a one-dimensional chain of quantum
dots~\cite{loss:qdot-comp}, the original Kane
proposal~\cite{kane:nature-si-qc}, and the all-silicon NMR
device~\cite{ladd:si-nmr-qc}.  Superconducting
qubits~\cite{pashkin:oscill,you:scalable} may map to $NTC$, depending
on the details of the qubit interconnection.
% Small-scale ion
% traps~\cite{cirac95:_cold_trapped_ions} and 

The difference between $AC$ and $NTC$ is critical; beyond the
important constant factors as nearby qubits shuffle, we will see in
section~\ref{sec:csla} that $AC$ can achieve $O(\log{n})$ performance
where $NTC$ is limited to $O(n)$.

\begin{figure}
%\def\epsfsize#1#2{.4\hsize}
%\centerline{\hbox{
%\epsfbox{fivegateccnot.eps}}}
\includegraphics[height=2cm]{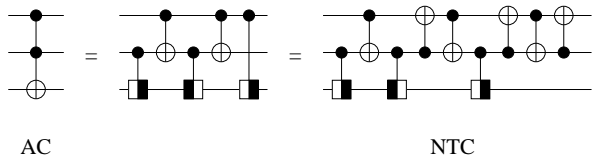}
\caption{CCNOT constructions for our architectures $AC$ and $NTC$.
  The box with the bar on the right represents the square root of $X$,
  and the box with the bar on the left its adjoint.  Time flows left
  to right, each horizontal line represents a qubit, and each vertical
  line segment is a quantum gate.}
\label{fig:5gcc}
\end{figure}

For $NTC$, which does not support CCNOT directly, we compose CCNOT
from a set of five two-qubit gates~\cite{barenco:elementary}, as shown
in figure~\ref{fig:5gcc}.  The box with the bar on the right
represents the square root of $X$,
$\sqrt{X} = \frac{1}{2}\left[\begin{array}{cccc} 1+i & 1-i \\ 1-i & 1+i
\end{array}\right]$
and the box with the bar on the left its adjoint.  We assume that this
gate requires the same execution time as a CNOT.

% covers concurrent, or parallel, quantum computation
Section~\ref{sec:basics} reviews Shor's algorithm and the need for
modular exponentiation, then summarizes the techniques we employ to
accelerate modular exponentiation.  The next subsection introduces the
best-known existing modular exponentiation algorithms and several
different adders. Section~\ref{sec:faster} begins by examining
concurrency in the lowest level elements, the adders.  This is
followed by faster adders and additional techniques for accelerating
modulo operations and exponentiation.  Section~\ref{sec:example} shows
how to balance these techniques and apply them to a specific
architecture and set of constraints.  We evaluate several complete
algorithms for our architectural models.  Specific gate latency
counts, rather than asymptotic values, are given for 128 bits and
smaller numbers.

% An appendix discusses larger multiplications in more detail.

\section{Basic Concepts}
\label{sec:basics}

\subsection{Modular Exponentiation and Shor's Algorithm}

Shor's algorithm for factoring numbers on a quantum computer uses the
quantum Fourier transform to find the order $r$ of a randomly chosen
number $x$ in the multiplicative group $(\bmod N)$.  This is achieved
by exponentiating $x$, modulo $N$, for a superposition of all possible
exponents $a$.  Therefore, efficient arithmetic algorithms to
calculate modular exponentiation in the quantum domain are critical.

Quantum modular exponentiation is the evolution of the state of a
quantum computer to hold 
\begin{equation}
|\psi\rangle|0\rangle\rightarrow
|\psi\rangle|x^\psi \bmod N\rangle
\end{equation}

When $|\psi\rangle$ is the superposition of all input states $a$ up to
a particular value $2N^2$,
\begin{equation}
|\psi\rangle = \frac{1}{N\sqrt{2}}\sum_{a=0}^{2N^2}|a\rangle
\end{equation}

The result is the superposition of the modular exponentiation of those
input states,
\begin{equation}
\frac{1}{N\sqrt{2}}\sum_{a=0}^{2N^2}|a\rangle|0\rangle\rightarrow
\frac{1}{N\sqrt{2}}\sum_{a=0}^{2N^2}|a\rangle|x^a \bmod N\rangle
\end{equation}

Depending on the algorithm chosen for modular exponentiation, $x$ may
appear explicitly in a register in the quantum computer, or may appear
only implicitly in the choice of instructions to be executed.

In general, quantum modular exponentiation algorithms are created from
building blocks that do modular multiplication,
\begin{equation}
|\alpha\rangle|0\rangle \rightarrow |\alpha\rangle|\alpha\beta \bmod
 N\rangle
\end{equation}
where $\beta$ and $N$ may or may not appear explicitly in quantum
registers.  This modular multiplication is built from blocks that
perform modular addition,
\begin{equation}
|\alpha\rangle|0\rangle \rightarrow |\alpha\rangle|\alpha + \beta \bmod
 N\rangle
\end{equation}
which, in turn, are usually built from blocks that perform addition
and comparison.

Addition of two $n$-bit numbers requires $O(n)$ gates.  Multiplication
of two $n$-bit numbers (including modular multiplication) combines the
convolution partial products (the one-bit products) of each pair of
bits from the two arguments.  This requires $O(n)$ additions of
$n$-bit numbers, giving a gate count of $O(n^2)$.  Our exponentiation
for Shor's algorithm requires $2n$ multiplications, giving a total
cost of $O(n^3)$.

Many of these steps can be conducted in parallel; in classical
computer system design, the {\em latency} or {\em circuit depth}, the
time from the input of values until the output becomes available, is
as important as the total computational complexity.  {\em Concurrency}
is the execution of more than one gate during the same execution time
slot.  We will refer to the number of gates executing in a time slot
as the concurrency or the concurrency level.  Our goal through the
rest of the paper is to exploit parallelism, or concurrency, to
shorten the total wall clock time to execute modular exponentiation,
and hence Shor's algorithm.

The algorithms as described here run on logical qubits, which will be
encoded onto physical qubits using quantum error correction
(QEC)~\cite{shor:qecc}.  Error correction processes are generally
assumed to be applied in parallel across the entire machine.
Executing gates on the encoded qubits, in some cases, requires
additional ancillae, so multiple concurrent logical gates will require
growth in physical qubit storage
space~\cite{steane03:ft-css-codes,steane02:_quant-entropy-arch}.
Thus, both physical and logical concurrency are important; in this
paper we consider only logical concurrency.

\subsection{Notation and Techniques for Speeding Up Modular
  Exponentation}
\label{sec:notation}
\label{sec:techniques}

In this paper, we will use $N$ as the number to be factored, and $n$
to represent its length in bits.  For convenience, we will assume that
$n$ is a power of two, and the high bit of $N$ is one.  $x$ is the
random value, smaller than $N$, to be exponentiated, and $|a\rangle$
is our superposition of exponents, with $a < 2N^2$ so that the length
of $a$ is $2n+1$ bits.

When discussing circuit cost, the notation is $(CCNOTs; CNOTs; NOTs)$
or $(CNOTs; NOTs)$.  The values may be total gates or circuit depth
(latency), depending on context.  The notation is sometimes enhanced
to show required concurrency and space, $(CCNOTs; CNOTs; NOTs)
\#(concurrency; space)$.

$t$ is time, or latency to execute an algorithm, and $S$ is space,
subscripted with the name of the algorithm or circuit subroutine.
When $t$ or $S$ is superscripted with $AC$ or $NTC$, the values are
for the latency of the construct on that architecture.  Equations
without superscripts are for an abstract machine assuming no
concurrency, equivalent to a total gate count for the $AC$
architecture.  $R$ is the number of calls to a subroutine, subscripted
with the name of the routine.

$m$, $g$, $f$, $p$, $b$, and $s$ are parameters that determine the
behavior of portions of our modular exponentiation algorithm.  $m$,
$g$, and $f$ are part of our carry-select/conditional-sum adder
(sec.~\ref{sec:csla}).  $p$ and $b$ are used in our indirection scheme
(sec.~\ref{sec:indirection}).  $s$ is the number of multiplier blocks
we can fit into a chosen amount of space (sec.~\ref{sec:conc-exp}).

% \subsection{Techniques for Speeding Up Modular Exponentiation}
% \label{sec:techniques}

Here we summarize the techniques which are detailed in following
subsections.  Our fast modular exponentiation circuit is built using
the following optimizations:

% nakahara04:qalg-accel,
\begin{itemize}
\item Select correct qubit layout and subsequences to implement gates,
  then hand optimize (no
  penalty)~\cite{vandersypen:thesis,aho:palindromes,kawano04:compiler,kunihiro:new-req,takahashi:_const_depth,yao93quantum,ahokas:_qft_complexity}.
\item Look for concurrency within addition/multiplication (no space
  penalty, maybe noise penalty)
  (secs.~\ref{sec:conc-vedral}).
\item Select multiplicand using table/indirection (exponential
  classical cost, linear reduction in quantum gate
  count)(\cite{van-meter:pay-the-exponential}, sec.~\ref{sec:indirection}).
\item Do multiplications concurrently (linear speedup for small
  values, linear cost in space, small gate count increase; requires
  quantum-quantum (Q-Q) multiplier, as well as classical-quantum (C-Q)
  multiplier) (sec.~\ref{sec:conc-exp}).
\item Move to e.g. carry-save adders ($n^2$ space penalty for
  reduction to log time, increases total gate
  count)(\cite{gossett98:q-carry-save}, sec.~\ref{sec:gossett})
  conditional-sum adders (sec.~\ref{sec:csum}), or carry-lookahead
  adders (sec.~\ref{sec:qcla}).
\item Reduce modulo comparisons, only do subtract $N$ on overflow
  (small space penalty, linear reduction in modulo arithmetic cost)
  (sec.~\ref{sec:cut-mod}).
\end{itemize}

% Also need to design for robustness.

\subsection{Existing Algorithms}

In this section we will review various components of the modular
exponentiation which will be used to construct our parallelized
version of the algorithm in section~\ref{sec:faster}.  There are many
ways of building adders and multipliers, and choosing the correct one
is a technology-dependent exercise~\cite{ercegovac-lang:dig-arith}.
Only a few classical techniques have been explored for quantum
computation.  The two most commonly cited modular exponentiation
algorithms are those of Vedral, Barenco, and
Ekert~\cite{vedral:quant-arith}, which we will refer to as VBE, and
Beckman, Chari, Devabhaktuni, and
Preskill~\cite{beckman96:eff-net-quant-fact}, which we will refer to
as BCDP.  Both BCDP and VBE algorithms build multipliers from variants
of carry-ripple adders, the simplest but slowest method; Cuccaro {\em
et al.} have recently shown the design of a smaller, faster
carry-ripple adder.  Zalka proposed a carry-select adder; we present
our design for such an adder in detail in section~\ref{sec:csla}.
Draper {\em et al.} have recently proposed a carry-lookahead adder,
and Gossett a carry-save adder.  Beauregard has proposed a circuit
that operates primarily in the Fourier transform space.

Carry-lookahead (sec.~\ref{sec:qcla}), conditional-sum
(sec.~\ref{sec:csum}), and carry-save (sec.~\ref{sec:gossett}) all
reach $O(\log n)$ performance for addition.  Carry-lookahead and
conditional-sum use more space than carry-ripple, but much less than
carry-save.  However, carry-save adders can be combined into fast
multipliers more easily.  We will see in sec.~\ref{sec:faster} how to
combine carry-lookahead and conditional-sum into the overall
exponentiation algorithms.

\subsubsection{VBE Carry-Ripple}

The VBE algorithm~\cite{vedral:quant-arith} builds full modular
exponentiation from smaller building blocks.  The bulk of the time is
spent in $20n^2-5n$ ADDERs~\footnote{When we write ADDER in all
capital letters, we mean the complete VBE $n$-bit construction, with
the necessary undo; when we write adder in small letters, we are
usually referring to a smaller or generic circuit block.}.  The full
circuit requires $7n+1$ qubits of storage: $2n+1$ for $a$, $n$ for the
other multiplicand, $n$ for a running sum, $n$ for the convolution
products, $n$ for a copy of $N$, and $n$ for carries.

In this algorithm, the values to be added in, the convolution partial
products of $x^a$, are programmed into a temporary register (combined
with a superposition of $|0\rangle$ as necessary) based on a control
line and a data bit via appropriate CCNOT gates.  The latency of ADDER
and the complete algorithm are
\begin{equation}
\label{eq:adder}
t_{ADD} = (4n-4; 4n-3; 0)
\end{equation}

%x^a \bmod N 
\begin{eqnarray}
\label{eq:vbe}
t_{V} &=& (20n^2-5n)t_{ADD}\nonumber\\
&=& (80n^3-100n^2+20n;\thinspace{}96n^3-84n^2+15n;  \nonumber\\
&& \thinspace{}8n^2-2n+1)
\end{eqnarray}

\subsubsection{BCDP Carry-Ripple}

The BCDP algorithm is also based on a carry-ripple adder.  It differs
from VBE in that it more aggressively takes advantage of classical
computation.  However, for our purposes, this makes it harder to use
some of the optimization techniques presented here.  Beckman {\em et
al.} present several optimizations and tradeoffs of space and time,
slightly complicating the analysis.

The exact sequence of gates to be applied is also dependent on the
input values of $N$ and $x$, making it less suitable for hardware
implementation with fixed gates (e.g., in an optical system).  In the
form we analyze, it requires $5n+3$ qubits, including $2n+1$ for
$|a\rangle$.  Borrowing from their equation 6.23,
\begin{eqnarray}
t_{B} &=& (54n^3-127n^2+108n-29; \nonumber\\
&&\thinspace{}10n^3+15n^2-38n+14;\nonumber\\
&&\thinspace{}20n^3-38n^2+22n-4)
\end{eqnarray}

% \section{Finding Optimizations and Concurrency in Existing Algorithms}

\subsubsection{Cuccaro Carry-Ripple}

Cuccaro {\em et al.} have recently introduced a carry-ripple circuit,
which we will call $CUCA$, which uses only a single ancilla
qubit~\cite{cuccaro04:new-quant-ripple}.  The latency of their adder
is $(2n-1;\thinspace{}5;\thinspace{}0)$ for the $AC$ architecture.

The authors do not present a complete modular exponentiation circuit;
we will use their adder in our algorithms {\bf F} and {\bf G}.  This
adder, we will see in section~\ref{sec:algo-g}, is the most efficient
known for $NTC$ architectures.

\subsubsection{Gossett Carry-Save and Carry-Ripple}
\label{sec:gossett}

Gossett's arithmetic is pure quantum, as opposed to the mixed
classical-quantum of BCDP.  Gossett does not provide a full modular
exponentiation circuit, only adders, multipliers, and a modular adder
based on the important classical techniques of {\em carry-save
arithmetic}~\cite{gossett98:q-carry-save}.

Gossett's carry-save adder, the important contribution of the paper,
can run in $O(\log n)$ time on $AC$ architectures.  It will remain
impractical for the foreseeable future due to the large number of
qubits required; Gossett estimates $8n^2$ qubits for a full
multiplier, which would run in $O(\log^2 n)$ time.  It bears further
analysis because of its high speed and resemblance to standard fast
classical multipliers.

Unfortunately, the paper's second contribution, Gossett's carry-ripple
adder, as drawn in his figure 7, seems to be incorrect.  Once fixed,
his circuit optimizes to be similar to VBE.

\subsubsection{Carry-Lookahead}
\label{sec:qcla}

Draper, Kutin, Rains, and Svore have recently proposed a
carry-lookahead adder, which we call
QCLA~\cite{draper04:quant-carry-lookahead}.  This method allows the
latency of an adder to drop to $O(\log n)$ for $AC$ architectures.
The latency and storage of their adder is
\begin{eqnarray}
t_{LA}^{AC} &=& (4\log_2 n + 3;\thinspace{}4;\thinspace{}2)\nonumber\\
&& \#(n;\thinspace4n-\log n-1)
\end{eqnarray}

The authors do not present a complete modular exponentiation circuit;
we will use their adder in our algorithm {\bf E}, which we evaluate
only for $AC$.  The large distances between gate operands make it
appear that QCLA is unattractive for $NTC$.

\subsubsection{Beauregard/Draper QFT-based Exponentiation}

Beauregard has designed a circuit for doing modular exponentiation in
only $2n+3$ qubits of space~\cite{beauregard03:small-shor}, based on
Draper's clever method for doing addition on Fourier-transformed
representations of numbers~\cite{draper00:quant-addition}.

The depth of Beauregard's circuit is $O(n^3)$, the same as VBE and
BCDP.  However, we believe the constant factors on this circuit are
very large; every modulo addition consists of four Fourier transforms
and five Fourier additions.

Fowler, Devitt, and Hollenberg have simulated Shor's algorithm using
Beauregard's algorithm, for a class of machine they call {\em linear
nearest neighbor}
($LNN$)~\cite{fowler04:_shor_implem,devitt04:_shor_qec_simul}.  $LNN$
corresponds approximately to our $NTC$.  In their implementation of
the algorithm, they found no significant change in the computational
complexity of the algorithm on $LNN$ or an $AC$-like abstract
architecture, suggesting that the performance of Draper's adder, like
a carry-ripple adder, is essentially architecture-independent.

\section{Results: Algorithmic Optimizations}
\label{sec:faster}
% \section{Faster Adders}

We present our concurrent variant of VBE, then move to
faster adders.  This is followed by methods for performing
exponentiation concurrently, improving the modulo arithmetic, and
indirection to reduce the number of quantum multiplications.

\subsection{Concurrent VBE}
\label{sec:conc-vedral}

\begin{figure}
%\def\epsfsize#1#2{.7\hsize}
%\centerline{\hbox{
%\epsfbox{concurrent-vedral-adder.eps}}}
\includegraphics{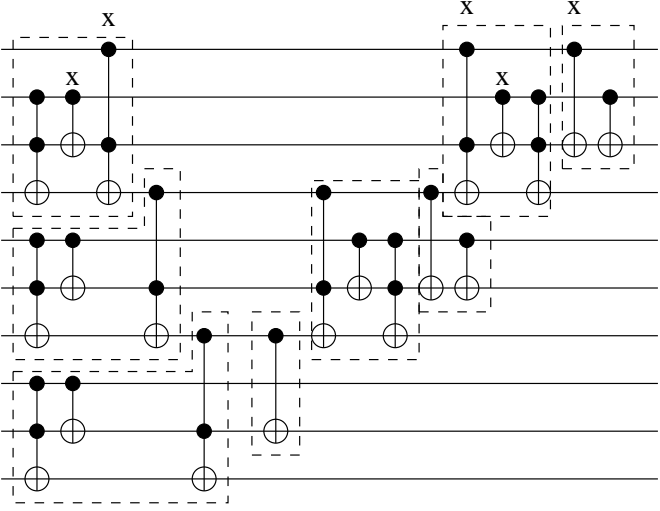}
\caption{Three-bit concurrent VBE ADDER, $AC$ abstract machine.
  Gates marked with an 'x' can be deleted when the carry in is known
  to be zero.}
\label{fig:cva}
\end{figure}

In figure~\ref{fig:cva}, we show a three-bit concurrent version of the
VBE ADDER.  This figure shows that the delay of the concurrent ADDER
is $(3n-3)CCNOT + (2n-3)CNOT$, or
\begin{equation}
t_{ADD}^{AC} = (3n-3;\thinspace{}2n-3;\thinspace{}0)
\label{eq:adderac}
\end{equation}
a mere 25\% reduction in latency compared to the unoptimized $(4n-4;
4n-3; 0)$ of equation~\ref{eq:adder}.

Adapting equation~\ref{eq:vbe}, the total circuit latency, minus a
few small corrections that fall outside the ADDER block proper, is
\begin{eqnarray}
t_{V}^{AC} &=& (20n^2-5n)t_{ADD}^{AC} \nonumber\\
&=& (60n^3-75n^2+15n; \nonumber\\
&& \thinspace{}40n^3-70n^2+15n;\thinspace{}0)
\end{eqnarray}
This equation is used to create the first entry in
table~\ref{tab:128-gate-count}.

\subsection{Carry-Select and Conditional-Sum Adders}
\label{sec:csla}

Carry-select adders concurrently calculate possible results without
knowing the value of the carry in.  Once the carry in becomes
available, the correct output value is selected using a multiplexer
(MUX).  The type of MUX determines whether the behavior is 
$O(\sqrt{n})$ or $O(\log n)$.

\subsubsection{$O(\sqrt{n})$ Carry-Select Adder}

The bits are divided into $g$ groups of $m$ bits each, $n = gm$.  The
adder block we will call CSLA, and the combined adder, MUXes, and
adder undo to clean our ancillae, CSLAMU.  The CSLAs are all executed
concurrently, then the output MUXes are cascaded, as shown in
figure~\ref{fig:csel4b}.  The first group may have a different size,
$f$, than $m$, since it will be faster, but for the moment we assume
they are the same.

\begin{figure}
%\def\epsfsize#1#2{.9\hsize}
%\centerline{\hbox{
%\epsfbox{two-bit-carry-select-adder.eps}}}
\includegraphics[width=8.6cm]{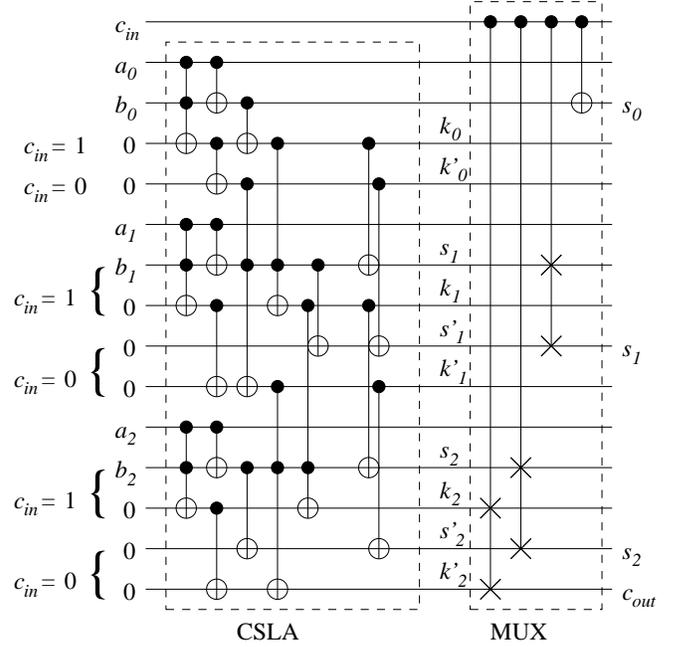}
\caption{Three-bit carry-select adder (CSLA) with multiplexer (MUX).
  $a_i$ and $b_i$ are addends.  The control-SWAP gates in the MUX
  select either the qubits marked $c_{in} = 1$ or $c_{in} = 0$
  depending on the state of the carry in qubit $c_{in}$.  $s_i$ qubits
  are the output sum and $k_i$ are internal carries.}
\label{fig:csel3}
\end{figure}

\begin{figure}
%\def\epsfsize#1#2{.75\hsize}
%\centerline{\hbox{
%\epsfbox{four-block-carry-select-adder.eps}}}
\includegraphics{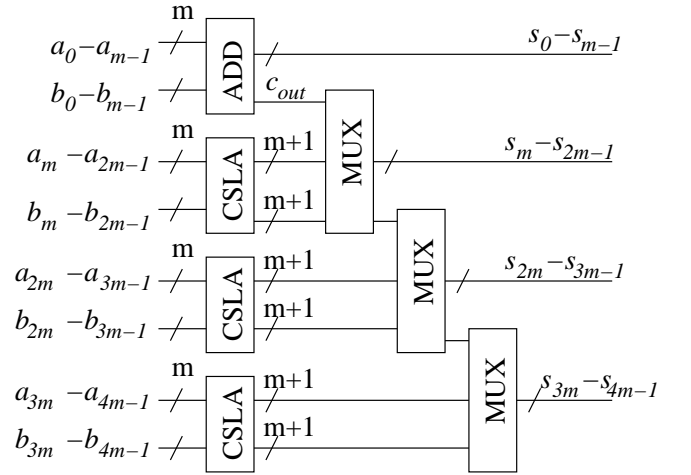}
\caption{Block-level diagram of four-group carry-select adder.  $a_i$
  and $b_i$ are addends and $s_i$ is the sum.  Additional ancillae not
  shown.}
\label{fig:csel4b}
\end{figure}

Figure~\ref{fig:csel3} shows a three-bit carry-select adder.  This
generates two possible results, assuming that the carry in will be
zero or one.  The portion on the right is a MUX used to select which
carry to use, based on the carry in.  All of the outputs without
labels are ancillae to be garbage collected.  It is possible that a
design optimized for space could reuse some of those qubits; as drawn
a full carry-select circuit requires $5m-1$ qubits to add two $m$-bit
numbers.

The larger $m$-bit carry-select adder can be constructed so that its
internal delay, as in a normal carry-ripple adder, is one additional
CCNOT for each bit, although the total number of gates increases and
the distance between gate operands increases.

The latency for the CSLA block is
\begin{equation}
\label{eq:cslaac}
t_{CS}^{AC} = (m;\thinspace{}2;\thinspace{}0)
\end{equation}
Note that this is not a ``clean'' adder; we still have ancillae to
return to the initial state.

The problem for implementation will be creating an efficient MUX,
especially on $NTC$.  Figure~\ref{fig:csel4b} makes it clear that the
total carry-select adder is only faster if the latency of MUX is
substantially less than the latency of the full carry-ripple.  It will
be difficult for this to be more efficient that the single-CCNOT delay
of the basic VBE carry-ripple adder on $NTC$.  On $AC$, it is
certainly easy to see how the MUX can use a fanout tree consisting of
more ancillae and CNOT gates to distribute the carry in signal, as
suggested by Moore~\cite{moore98:_parallel_quantum}, allowing all MUX
Fredkin gates to be executed concurrently.  A full fanout requires an
extra $m$ qubits in each adder.

In order to unwind the ancillae to reuse them, the simplest approach
is the use of CNOT gates to copy our result to another $n$-bit
register, then a reversal of the circuitry.  Counting the copy out for
ancilla management, we can simplify the MUX to two CCNOTs and
a pair of NOTs.

The latency of the carry ripple from MUX to MUX (not qubit to qubit)
can be arranged to give a MUX cost of $(4g+2m-6;0;2g-2)$.  This cost
can be accelerated somewhat by using a few extra qubits and ``fanning
out'' the carry.  For intermediate values of $m$, we will use a fanout
of 4 on $AC$, reducing the MUX latency to $(4g+m/2-6;2;2g-2)$ in
exchange for 3 extra qubits in each group.

Our space used for the full, clean adder is $(6m-1)(g-1)+3f+4g$ when
using a fanout of 4.

The total latency of the CSLA, MUX, and the CSLA undo is
\begin{eqnarray}
\label{eq:cslamuac}
t_{SEM}^{AC} &=& 2t_{CS}^{AC}+t_{MUX}^{AC} \nonumber\\
&=& (4g+5m/2-6;\thinspace{}6;\thinspace{}2g-2)
\end{eqnarray}
Optimizing for $AC$, based on equation~\ref{eq:cslamuac}, the delay
will be the minimum when $m \sim \sqrt{8n/5}$.

Zalka was the first to propose use of a carry-select adder, though he
did not refer to it by name~\cite{zalka98:_fast_shor}.  His analysis
does not include an exact circuit, and his results differ
slightly from ours.

\subsubsection{$O(\log n)$ Conditional Sum Adder}
\label{sec:csum}

\begin{figure}
\includegraphics[height=10cm]{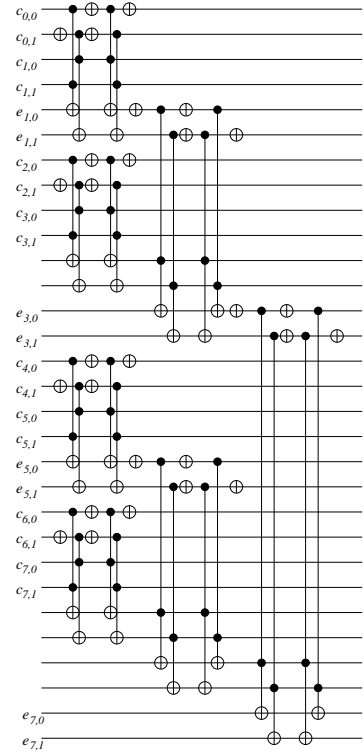}
\caption{$O(\log n)$ MUX for conditional-sum adder, for $g=9$ (the
  first group is not shown).  Only the $c_{i,j}$ carry out lines from
  each $m$-qubit block are shown, where $i$ is the block number and
  $j$ is the carry in value.  At each stage, the span of correct
  effective swap control lines $e_{i,j}$ doubles.  After using the
  swap control lines, all but the last must be cleaned by reversing
  the circuit.  Unlabeled lines are ancillae to be cleaned.}
\label{fig:csla-new-mux}
\end{figure}

As described above, the carry-select adder is $O(m + g)$, for $n =
mg$, which minimizes to be $O(\sqrt{n})$.  To reach $O(\log n)$
performance, we must add a multi-level MUX to our carry-select adder.
This structure is called a conditional sum adder, which we will label
CSUM.  Rather than repeatedly choosing bits at each level of the MUX,
we will create a multi-level distribution of MUX select signals, then
apply them once at the end.  Figure~\ref{fig:csla-new-mux} shows only
the carry signals for eight CSLA groups.  The $e$ signals in the
figure are our effective swap control signals.  They are combined with
a carry in signal to control the actual swap of variables.  In a full
circuit, a ninth group, the first group, will be a carry-ripple adder
and will create the carry in; that carry in will be distributed
concurrently in a separate tree.

The total adder latency will be
\begin{eqnarray}
t_{CSUM}^{AC} &=& 2t_{CS}^{AC} +\nonumber\\
&& (2\lceil \log_2(g-1)\rceil - 1)
\times(2;\thinspace{}0;\thinspace{}2)\nonumber\\
&& + (4;\thinspace{}0;\thinspace{}4)\nonumber\\
&=&(2m + 4\lceil\log_2 (g-1)\rceil + 2;\thinspace{}4;\nonumber\\
&&\thinspace{}4\lceil\log_2 (g-1)\rceil+2)
\end{eqnarray}
where $\lceil x\rceil$ indicates the smallest integer not smaller than
$x$.

For large $n$, this generally reaches a minimum for small $m$, which
gives asymptotic behavior $\sim4\log_2 n$, the same as QCLA.  CSUM
is noticeably faster for small $n$, but requires more space.

The MUX uses $\lceil 3(g-1)/2\rceil - 2$ qubits in addition to the
internal carries and the tree for dispersing the carry in.  Our space
used for the full, clean adder is $(6m-1)(g-1)+3f+\lceil 3(g-1)/2
-2+ (n-f)/2\rceil$.

%\section{Further Algorithmic Optimizations}

\subsection{Concurrent Exponentiation}

\label{sec:conc-exp}

Modular exponentiation is often drawn as a string of modular
multiplications, but Cleve and Watrous pointed out that these can
easily be parallelized, at linear cost in space~\cite{cleve:qft}.
We always have to execute $2n$ multiplications; the goal is to do
them in as few time-delays as possible.

\begin{figure}
%\def\epsfsize#1#2{.8\hsize}
%\centerline{\hbox{
% \input{pmodexp.pstex_t}}}
\includegraphics[width=8.6cm]{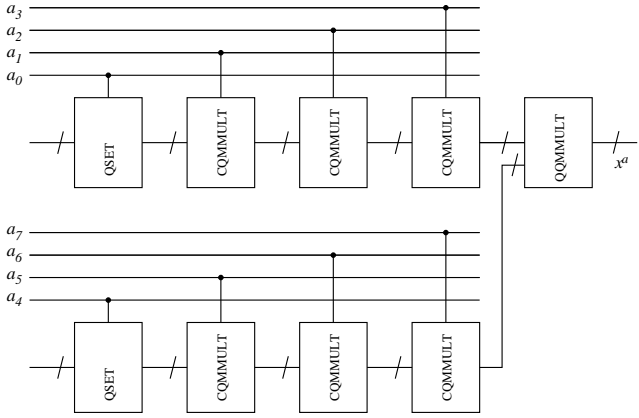}
\caption{Concurrent modular multiplication in modular exponentiation
  for $s = 2$. QSET simply sets the sum register to the appropriate
  value.}
\label{fig:pmodexp}
\end{figure}

To go (almost) twice as fast, use two multipliers.  For four times,
use four.  Naturally, this can be built up to $n$ multipliers to
multiply the necessary $2n+1$ numbers, in which case a tree
recombining the partial results requires $\log_2 n$ quantum-quantum
(Q-Q) multiplier latency times.  The first unit in each chain just
sets the register to the appropriate value if the control line is 1,
otherwise, it leaves it as 1.

For $s$ multipliers, $s \le n$, each multiplier must combine $r =
\lfloor (2n+1)/s \rfloor$ or $r+1$ numbers, using $r-1$ or $r$
multiplications (the first number being simply set into the running
product register), where $\lfloor x\rfloor$ indicates the largest
integer not larger than $x$.  The intermediate results from the
multipliers are combined using $\lceil \log_2 s\rceil$ Q-Q
multiplication steps.

For a parallel version of VBE, the exact latency, including cases
where $rs \neq 2n + 1$, is
\begin{eqnarray}
\label{eq:vmults}
R_{V} &=&
% \begin{cases}
% 2(t-1) + \lceil \log_2 s \rceil & ts = 2n \\
% 2t + \lceil \log_2 (\lceil(s- 2n + ts)/4\rceil + (2n - ts))\rceil & ts \neq 2n
2r  + 1 + \lceil \log_2 (\lceil(s- 2n - 1 + rs)/4\rceil\nonumber\\
&& + 2n + 1 - rs)\rceil
% \end{cases}
\end{eqnarray}
times the latency of our multiplier.  For small $s$, this is $O(n)$;
for larger $s$,
\begin{equation}
\lim_{s\rightarrow n} O(n/s + \log s) = O(\log n)
\end{equation}

\subsection{Reducing the Cost of Modulo Operations}
\label{sec:cut-mod}

The VBE algorithm does a trial subtraction of $N$ in each modulo
addition block; if that underflows, $N$ is added back in to the total.
This accounts for two of the five ADDER blocks and much of the extra
logic to compose a modulo adder.  The last two of the five blocks are
required to undo the overflow bit.

\begin{figure}
%\def\epsfsize#1#2{1.0\hsize}
%\centerline{\hbox{
%\epsfbox{eff-mod-add-2.eps}}}
\includegraphics[width=8.6cm]{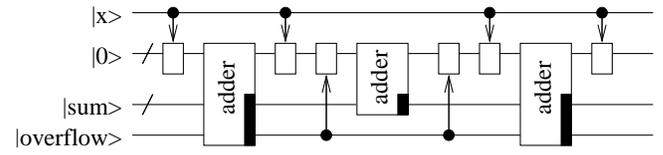}
\caption{More efficient modulo adder.  The blocks with arrows set the
  register contents based on the value of the control line.  The
  position of the black block indicates the running sum in our
  output.}
\label{fig:ema2}
\end{figure}

Figure~\ref{fig:ema2} shows a more efficient modulo adder than VBE,
based partly on ideas from BCDP and Gossett.  It requires only three
adder blocks, compared to five for VBE, to do one modulo addition.
The first adder adds $x^j$ to our running sum.  The second
conditionally adds $2^n-x^j-N$ or $2^n-x^j$, depending on the value of
the overflow bit, {\em without} affecting the overflow bit, arranging
it so that the third addition of $x^j$ will overflow and clear the
overflow bit if necessary.  The blocks pointed to by arrows are the
addend register, whose value is set depending on the control lines.
Figure~\ref{fig:ema2} uses $n$ fewer bits than VBE's modulo
arithmetic, as it does not require a register to hold $N$.

In a slightly different fashion, we can improve the performance of
VBE by adding a number of qubits, $p$, to our result register, and
postponing the modulo operation until later.  This works as long as we
don't allow the result register to overflow; we have a redundant
representation of modulo $N$ values, but that is not a problem at this
stage of the computation.

The largest number that doesn't overflow for $p$ extra qubits is
$2^{n+p}-1$; the largest number that doesn't result in subtraction is
$2^{n+p-1}-1$.  We want to guarantee that we always clear that
high-order bit, so if we subtract $bN$, the most iterations we can go
before the next subtraction is $b$.

The largest multiple of $N$ we can subtract is
$\lfloor2^{n+p-1}/N\rfloor$.  Since $2^{n-1} < N < 2^n$, the largest
$b$ we can allow is, in general, $2^{p-1}$.

For example, adding three qubits, $p = 3$, allows $b = 4$, reducing
the 20 ADDER calls VBE uses for four additions to 9 ADDER calls, a
55\% performance improvement.  As $p$ grows larger, the cost of the
adjustment at the end of the calculation also grows and the additional
gains are small.  We must use $3p$ adder calls at the end of the
calculation to perform our final modulo operation.  Calculations
suggest that $p$ of up to 10 or 11 is still faster.

The equation below shows the number of calls to our adder block
necessary to make an $n$-bit modulo multiplier.
\begin{equation}
R_{M} = n(2b+1)/b
\end{equation}

\subsection{Indirection}
\label{sec:indirection}

We have shown elsewhere that it is possible to build a table
containing small powers of $x$, from which an argument to a multiplier
is selected~\cite{van-meter:pay-the-exponential}.  In exchange for
adding storage space for $2^w$ $n$-bit entries in a table, we can
reduce the number of multiplications necessary by a factor of $w$.
This appears to be attractive for small values of $w$, such as 2 or 3.

\begin{figure}
%\def\epsfsize#1#2{1.0\hsize}
%\centerline{\hbox{
%\epsfbox{implicit-indirect.eps}}}
\includegraphics{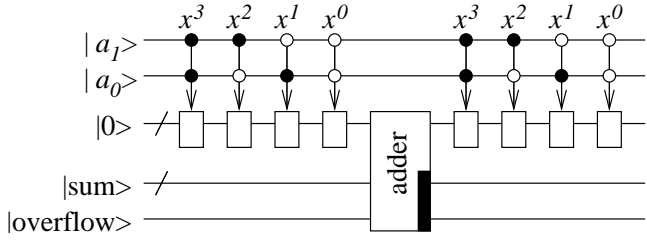}
\caption[width=8.6cm]{Implicit Indirection.  The arrows pointing to
  blocks indicate the setting of the addend register based on the
  control lines.  This sets the addend from a table stored in
  classical memory, reducing the number of quantum multiplications by
  a factor of $w$ in exchange for $2^w$ argument setting operations.}
\label{fig:imp-ind}
\end{figure}

In our prior work, we proposed using a large quantum memory, or a
quantum-addressable classical memory
(QACM)~\cite{nielsen-chuang:qci-qacm}.  Here we show that the quantum
storage space need not grow; we can implicitly perform the lookup by
choosing which gates to apply while setting the argument.  In
figure~\ref{fig:imp-ind}, we show the setting and resetting of the
argument for $w = 2$, where the arrows indicate CCNOTs to set the
appropriate bits of the 0 register to 1.  The actual implementation
can use a calculated enable bit to reduce the CCNOTs to CNOTs.  Only
one of the values $x^0$, $x^1$, $x^2$, or $x^3$ will be enabled, based
on the value of $|a_{1}a_{0}\rangle$.

The setting of this input register may require propagating $|a\rangle$
or the enable bit across the entire register.  Use
of a few extra qubits ($2^{w-1}$) will allow the several setting
operations to propagate in a tree.
\begin{equation}
t_{ARG}^{AC} = 
\begin{cases}
2^w(1;0;1) = (4; 0; 4)& w = 2 \\
2^w(3;0;1) & w = 3,4 \end{cases}
\end{equation}

\begin{figure}
%\def\epsfsize#1#2{\hsize}
%\centerline{\hbox{
%\epsfbox{two-gate-carry-all.eps}}}
\includegraphics[width=8.6cm]{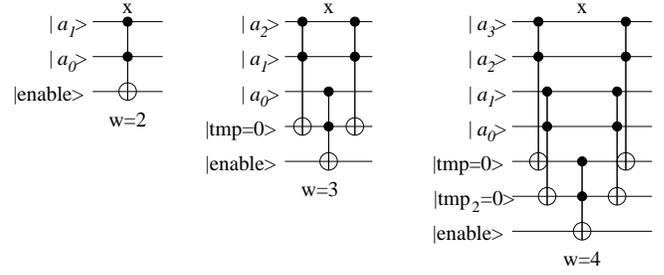}
\caption{Argument setting for indirection for different values of $w$,
for the $AC$ architecture.  For the $w=4$ case, the two CCNOTs on the
left can be executed concurrently, as can the two on the right, for a
total latency of 3.}
\label{fig:arg-setting}
\end{figure}

For $w = 2$ and $w = 3$, we calculate that setting the argument adds
$(4;0;4)\#(4,5)$ and $(24;0;8)\#(8,9)$, respectively, to the latency,
concurrency and storage of each adder.  We create separate
enable signals for each of the $2^w$ possible arguments and pipeline
flowing them across the register to set the addend bits.  We consider
this cost only when using indirection.  Figure~\ref{fig:arg-setting}
shows circuits for $w=2,3,4$.

Adapting equation~\ref{eq:vmults} to both indirection and concurrent
multiplication, we have a total latency for our circuit, in multiplier
calls, of
\begin{equation}
\label{eq:imults}
R_{I} = 
2r + 1 + \lceil \log_2 (\lceil(s- 2n - 1 + rs)/4\rceil + 2n + 1 - rs)\rceil
\end{equation}
where $r = \lfloor \lceil (2n+1)/w\rceil / s \rfloor$.

%\section{Optimizing for an Architecture}
%\section{Putting It All Together}
\section{Example: Exponentiating a 128-bit Number}
\label{sec:example}

%\subsection{Balancing Techniques}

In this section, we combine these techniques into complete algorithms
and examine the performance of modular exponentiation of a 128-bit
number.  We assume the primary engineering constraint is the available
number of qubits.  In section~\ref{sec:conc-exp} we showed that using
twice as much space can almost double our speed, essentially linearly
until the log term begins to kick in.  Thus, in managing space
tradeoffs, this will be our standard; any technique that raises
performance by more than a factor of $c$ in exchange for $c$ times as
much space will be used preferentially to parallel multiplication.
Carry-select adders (sec.~\ref{sec:csla}) easily meet this criterion,
being perhaps six times faster for less than twice the space.

Algorithm {\bf D} uses $100n$ space and our conditional-sum adder
$CSUM$.  Algorithm {\bf E} uses $100n$ space and the
carry-lookahead adder $QCLA$.  Algorithms {\bf F} and {\bf G} use the
Cuccaro adder and $100n$ and minimal space, respectively.  Parameters
for these algorithms are shown in table~\ref{tab:algorithms}.  We have
included detailed equations for concurrent VBE and {\bf D} and numeric
results in table~\ref{tab:128-gate-count}.  The performance ratios are
based only on the CCNOT gate count for $AC$, and only on the CNOT gate
count for $NTC$.

\begin{table*}
\begin{tabular}{|l|r|r|r|r|r|r|} \hline
algorithm & adder & modulo & indirect & multipliers ($s$) & space & concurrency \\ \hline
concurrent VBE & VBE & VBE & N/A & 1 & 897 & 2 \\
algorithm {\bf D}
& CSUM($m = 4$) & $p=11, b=1024$ & $w=2$ & 12 & 11969 & $126\times 12=1512$	\\
algorithm {\bf E}
& QCLA & $p=10, b=512$ & $w=2$ & 16 & 12657 & $128\times 16 = 2048$	\\
algorithm {\bf F}
& CUCA & $p=10, b=512$ & $w=4$ & 20 & 11077 & $20\times 2 = 40$	\\
algorithm {\bf G}
& CUCA & fig.~\ref{fig:ema2} & $w=4$ & 1 & 660 & $2$	\\
\hline
\end{tabular}
\caption{Parameters for our algorithms, chosen for 128 bits.}
\label{tab:algorithms}
\end{table*}

\begin{table*}
\begin{tabular}{|l|r|r|r|r|r|} \hline
algorithm	  	 & \multicolumn{2}{c|}{$AC$} & \multicolumn{2}{c|}{$NTC$} \\
 \hline
& gates & perf. & gates & perf. \\
\hline
concurrent VBE &
 $(1.25\times 10^{8};\thinspace{}8.27\times
 10^{7};\thinspace{}0.00\times 10^{0})$ & 1.0 &
 $(8.32\times 10^{8};\thinspace{}0.00\times 10^{0})$      & 1.0 \\
algorithm D             &
 $(2.19\times 10^{5};\thinspace{}2.57\times
 10^{4};\thinspace{}1.67\times 10^{5})$ & 569.8 &
 N/A    & N/A \\
algorithm E             &
 $(1.71\times 10^{5};\thinspace{}1.96\times
 10^{4};\thinspace{}2.93\times 10^{4})$ & 727.2 &
 N/A    & N/A \\
algorithm F             &
 $(7.84\times 10^{5};\thinspace{}1.30\times
 10^{4};\thinspace{}4.10\times 10^{4})$ & 158.9 &
 $(4.11\times 10^{6};\thinspace{}4.10\times 10^{4})$      & 202.5 \\
algorithm G             &
 $(1.50\times 10^{7};\thinspace{}2.48\times
 10^{5};\thinspace{}7.93\times 10^{5})$ & 8.3 &
 $(7.87\times 10^{7};\thinspace{}7.93\times 10^{5})$      & 10.6 \\
\hline
\end{tabular}
\caption{Latency to factor a 128-bit number for various architectures
  and choices of algorithm.  $AC$, abstract concurrent
  architecture. $NTC$ neighbor-only, two-qubit gate, concurrent
  architecture.  perf, performance relative to VBE algorithm for that
  architecture, based on CCNOTs for $AC$ and CNOTs for $NTC$.}
\label{tab:128-gate-count}
\end{table*}

\subsection{Concurrent VBE}

On $AC$, the concurrent VBE ADDER is $(3n-3; 2n-3; 0) = (381;253;0)$
for 128 bits.  This is the value we use in the concurrent VBE line in
table~\ref{tab:128-gate-count}.  This will serve as our best baseline
time for comparing the effectiveness of more drastic algorithmic
surgery.

\begin{figure*}
%\def\epsfsize#1#2{\hsize}
%\centerline{\hbox{
%\epsfbox{nmr-three-bit-vedral-adder.eps}}}
\includegraphics[width=17.2cm]{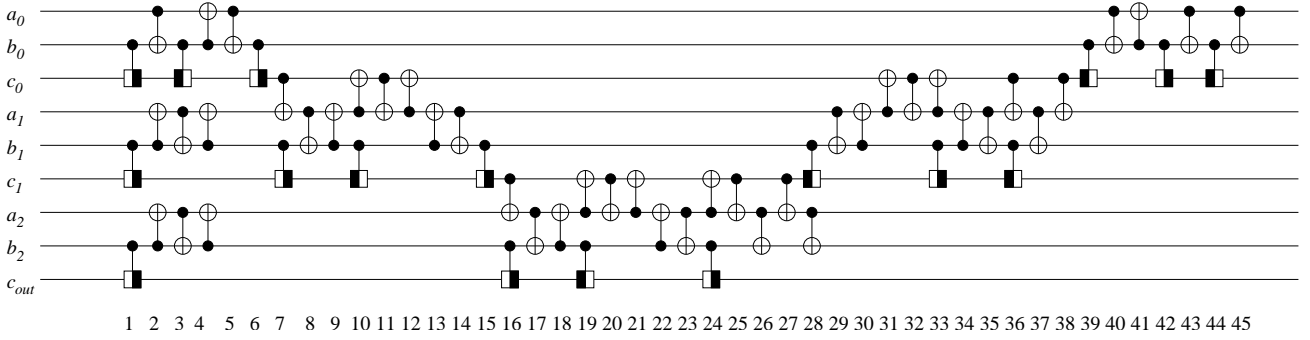}
\caption{Optimized, concurrent three bit VBE ADDER for the $NTC$
  architecture.  Numbers across the bottom are time steps.}
\label{fig:nmr-adder}
\end{figure*}

Figure~\ref{fig:nmr-adder} shows a fully optimized, concurrent, but
otherwise unmodified version of the VBE ADDER for three bits on a
neighbor-only machine ($NTC$ architecture), with the gates marked 'x'
in figure~\ref{fig:cva} eliminated.  The latency is 
\begin{equation}
t_{ADD}^{NTC} = (20n-15;0)\#(2;\thinspace{}3n+1)
\end{equation}
or 45 gate times for the three-bit adder.  A 128-bit adder will have a
latency of $(2545;0)$.  The diagram shows a concurrency level of
three, but simple adjustment of execution time slots can limit that to
two for any $n$, with no latency penalty.

The unmodified full VBE modular exponentiation algorithm, consists
of $20n^2-5n = 327040$ ADDER calls plus minor additional logic.

\begin{eqnarray}
t_{V}^{NTC} &=& (20n^2-5n)t_{ADD}^{NTC} \nonumber\\
&=& (400n^3-400n^2+75n;\thinspace{}0)
\end{eqnarray}

\subsection{Algorithm {\bf D}}

The overall structure of algorithm {\bf D} is similar to VBE, with our
conditional-sum adders instead of the VBE carry-ripple, and our
improvements in indirection and modulo.  As we do not consider CSUM
to be a good candidate for an algorithm for $NTC$, we evaluate only
for $AC$.  Algorithm {\bf D} is the fastest algorithm for $n=8$ and
$n=16$.

\begin{eqnarray}
t_{D} &=& R_{I} R_{M}\nonumber\\
&& \times(t_{CSUM}+t_{ARG})\nonumber\\
&& + 3pt_{CSUM}
\end{eqnarray}

Letting $r = \lfloor \lceil (2n+1)/w\rceil / s \rfloor$, the latency
and space requirements for algorithm {\bf D} are
\begin{eqnarray}
t_{D}^{AC} &=& 2r + 1 + \lceil \log_2 (\lceil(s- 2n - 1 +
 rs)/4\rceil\nonumber\\
&& + 2n + 1 - rs)\rceil n(2b+1)/b\nonumber\\
&& \times ((2m + 4\lceil\log_2 (g-1)\rceil + 2;\thinspace{}4;\nonumber\\
&&\thinspace{}4\lceil\log_2 (g-1)\rceil+2)
 +(4;\thinspace{}0;\thinspace{}4))\nonumber\\
&& + 3p(2m + 4\lceil\log_2 (g-1)\rceil + 2;\thinspace{}4;\nonumber\\
&&\thinspace{}4\lceil\log_2 (g-1)\rceil+2)
\end{eqnarray}
and
\begin{eqnarray}
S_{D} &=& s(S_{CSUM}\nonumber\\
&& + 2^w + 1 + p + n) + 2n + 1\nonumber\\
&=& s(7n - 3m - g + 2^w + p\nonumber\\
&& + \lceil3(g-1)/2 - 2 + (n-m)/2\rceil)\nonumber\\
&& + 2n + 1
\end{eqnarray}

\subsection{Algorithm {\bf E}}

Algorithm {\bf E} uses the carry-lookahead adder QCLA in place of the
conditional-sum adder CSUM.  Although CSUM is slightly faster
than QCLA, its significantly larger space consumption means that in
our $100n$ fixed-space analysis, we can fit in 16 multipliers using
QCLA, compared to only 12 using CSUM.  This allows the overall
algorithm {\bf E} to be 28\% faster than {\bf D} for 128 bits.

\subsubsection{Algorithms {\bf F} and {\bf G}}
\label{sec:algo-g}

The Cuccaro carry-rippler adder has a latency of
$(10n+5;\thinspace{}0)$ for $NTC$.  This is twice as fast as the VBE
adder.  We use this in our algorithms {\bf F} and {\bf G}.  Algorithm
{\bf F} uses $100n$ space, while {\bf G} is our attempt to produce the
fastest algorithm in the minimum space.

\subsection{Smaller $n$ and Different Space}

\begin{figure}
%\def\epsfsize#1#2{\hsize}
%\centerline{\hbox{
%\epsfbox{two-gate-carry-all.eps}}}
\includegraphics[width=8.6cm]{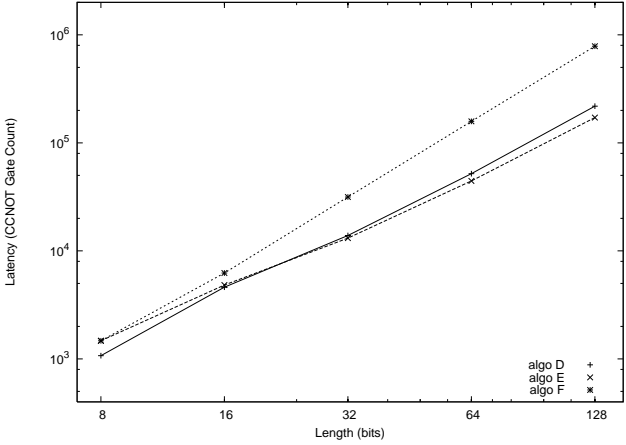}
\caption{Execution time for our algorithms for space $100n$ on the
  $AC$ architecture, for varying value of $n$.}
\label{fig:exp-graph}
\end{figure}

Figure~\ref{fig:exp-graph} shows the execution times of our three
fastest algorithms for $n$ from eight to 128 bits.  Algorithm {\bf D},
using CSUM, is the fastest for eight and 16 bits, while {\bf E},
using QCLA, is fastest for larger values.  The latency of 1072 for $n
= 8$ bits is 32 times faster than concurrent VBE, achieved with $60n =
480$ qubits of space.

\begin{figure}
\includegraphics[width=8.6cm]{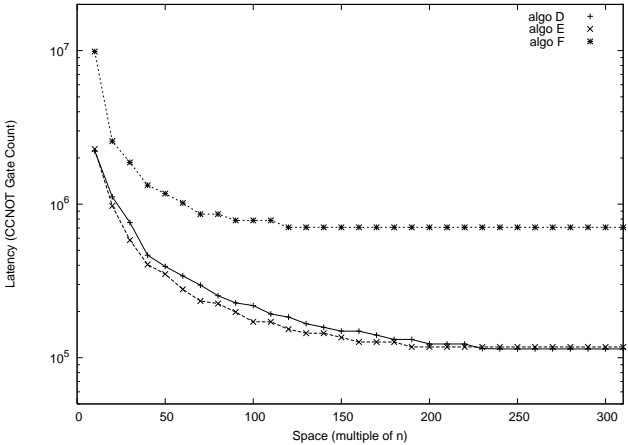}
\caption{Execution time for our algorithms for 128 bits on the
  $AC$ architecture, for varying multiples of $n$ space available.}
\label{fig:space-graph}
\end{figure}

Figure~\ref{fig:space-graph} shows the execution times for $n = 128$
bits for various amounts of available space.  All of our algorithms
have reached a minimum by $240n$ space (roughly $1.9n^2$).

\subsection{Asymptotic Behavior}

The focus of this paper is the constant factors in modular
exponentiation for important problem sizes and architectural
characteristics.  However, let us look briefly at the asymptotic
behavior of our circuit depth.

In section~\ref{sec:conc-exp}, we showed that the latency of our
complete algorithm is
\begin{equation}
\label{eq:expcost}
 O(n/s + \log s)\times\text{latency of multiplication}
\end{equation}
as we parallelize the multiplication using $s$ multiplier blocks.  Our
multiplication algorithm is still
\begin{equation}
\label{eq:multcost}
 O(n)\times\text{latency of addition}
\end{equation}

Algorithms {\bf D} and {\bf E} both use an $O(\log n)$ adder.
Combining equations \ref{eq:expcost} and \ref{eq:multcost} with the
adder cost, we have asymptotic circuit depth of
\begin{equation}
t_{D}^{AC} = t_{E}^{AC} = O((n\log n)(n/s + \log s))
\end{equation}
 for algorithms {\bf D} and {\bf E}.  As $s\rightarrow n$, these
approach $O(n \log^2 n)$ and space consumed approaches $O(n^2)$.

Algorithm {\bf F} uses an $O(n)$ adder, whose asymptotic behavior is
the same on both $AC$ and $NTC$, giving
\begin{equation}
t_{F}^{AC} = t_{F}^{NTC} = O((n^2)(n/s + \log s))
\end{equation}
approaching $O(n^2 \log n)$ as space consumed approaches $O(n^2)$.

This compares to asymptotic behavior of $O(n^3)$ for VBE, BCDP, and
algorithm {\bf G}, using $O(n)$ space.  The limit of performance,
using a carry-save multiplier and large $s$, will be $O(\log^3 n)$ in
$O(n^3)$ space.

\section{Discussion and Future Work}

We have shown that it is possible to significantly accelerate quantum
modular exponentiation using a stable of techniques.  We have provided
exact gate counts, rather than asymptotic behavior, for the $n = 128$
case, showing algorithms that are faster by a factor of 200 to 700,
depending on architectural features, when $100n$ qubits of storage are
available.  For $n = 1024$, this advantage grows to more than a factor
of 5,000 for non-neighbor machines ($AC$).  Neighbor-only ($NTC$)
machines can run algorithms such as addition in $O(n)$ time at best,
when non-neighbor machines ($AC$) can achieve $O(\log{n})$
performance.

In this work, our contribution has focused on parallelizing execution
of the arithmetic through improved adders, concurrent gate execution,
and overall algorithmic structure.  We have also made improvements
that resulted in the reduction of modulo operations, and traded some
classical for quantum computation to reduce the number of quantum
operations.  It seems likely that further improvements can be found in
the overall structure and by more closely examining the construction
of multipliers from adders~\cite{ercegovac-lang:dig-arith}.  We also
intend to pursue multipliers built from hybrid carry-save adders.

The three factors which most heavily influence performance of modular
exponentiation are, in order, concurrency, the availability of large
numbers of application-level qubits, and the topology of the
interconnection between qubits.  Without concurrency, it is of course
impossible to parallelize the execution of any algorithm.  Our
algorithms can use up to $\sim2n^2$ application-level qubits to
execute the multiplications in parallel, executing $O(n)$
multiplications in $O(\log n)$ time steps.  Finally, if any two qubits
can be operands to a quantum gate, regardless of location, the
propagation of information about the carry allows an addition to be
completed in $O(\log n)$ time steps instead of $O(n)$.  We expect that
these three factors will influence the performance of other algorithms
in similar fashion.

Not all physically realizable architectures map cleanly to one of our
models.  A full two-dimensional mesh, such as neutral atoms in an
optical lattice~\cite{brennen99:_optical_lattice}, and a loose trellis
topology~\cite{oskin:quantum-wires} probably fall between $AC$ and
$NTC$.  The behavior of the scalable ion
trap~\cite{kielpinski:large-scale} is not immediately clear.  We have
begun work on expanding our model definitions, as well as additional
ways to characterize quantum computer architectures.

The process of designing a large-scale quantum computer has only just
begun.  Over the coming years, we expect advances in the fundamental
technology, the system architecture, algorithms, and tools such as
compilers to all contribute to the creation of viable quantum
computing machines.  Our hope is that the algorithms and techniques in
this paper will contribute to that engineering process in both the
short and long term.

\section*{Acknowledgments}

The authors would like to thank Eisuke Abe, Fumiko Yamaguchi, and
Kevin Binkley of Keio University, Thaddeus Ladd of Stanford
University, Seth Lloyd of MIT, Y. Kawano and Y. Takahashi of NTT Basic
Research Laboratories, Bill Munro of HP Labs, Kae Nemoto of NII, and
the referee for helpful discussions and feedback.

\end{document}